\documentstyle[preprint,revtex,eqsecnum]{aps}
\begin{document}
\draft
\preprint{UAHEP-939}
\begin{title}
Proper Field Quantization in Black Hole Spacetimes
\end{title}
\author{B.Harms and Y.Leblanc}
\begin{instit}
Department of Physics and Astronomy, The University of
Alabama\\
Box 870324, Tuscaloosa, AL 35487-0324
\end{instit}
\begin{abstract}
Canonical quantization of local field theories in classical
black
hole spacetimes with a single horizon leads to a particle
number
density with a thermal distribution in equilibrium at the
Hawking
temperature. A complete treatment including non-local
quantum
gravity effects has shown however that the full ``thermal
vacuum''
of the theory is the false vacuum. In this work, we find the
true
vacuum consistent with the complete semiclassical analysis
of
quantum black holes. The theory is described by a
``microcanonical''
quantum field theory with fixed total energy $E=M$, the mass
of the
black hole. Considerations making use of the microcanonical
density
matrix as well as the idempotency condition show that
particles
in black hole backgrounds are described by pure states,
unlike the
canonical formulation.
\end{abstract}

\narrowtext
\section{Introduction}

Canonical quantization of local field theories in curved
spacetimes
and especially spacetimes with horizons (such as black
holes), was a
very popular subject in the late seventies and early
eighties.

In those years and until very recently, it was taken for a
fact of
life that the particle number density obtained in spacetimes
with a
single horizon  was given by a thermal distribution with a
temperature
equal to the Gibbons-Hawking temperature.

Of course, such a result meant the loss of quantum coherence
in those
systems and it was very popular to assume that indeed this
was the
case.

A WKB semiclassical analysis of quantum gravity however
suggested
otherwise, as quantum black holes were essentially
identifiable as
excitations of p-brane theories and were shown never to
achieve thermal
equilibrium.

In recent papers\cite{hl1,hl2,hl3,hl4,hl5}, and especially
the one of Ref.\cite{hl6},
the
present authors
presented a proof that the canonical
(thermal)
vacuum of local quantum field theories in black hole
spacetimes is
{\it not} the true vacuum of these theories. Crucial to the
demonstration was the incorporation of all the quantum
gravity (p-branes)
excitation modes (back reaction effects) in the construction
of
such a vacuum. This incorporation fully takes into account
the (quantum)
non-locality of quantum gravity theories. It was then shown
that the
stability of the full thermal vacuum was tantamount to the
existence
of black hole solutions to the so-called Hagedorn self-
consistency
condition.

Of course, it is well known that string theories are the
only solutions
to Hagedorn's condition and therefore, black hole solutions
are excluded.
The canonical vacuum is the false vacuum.

Our study of black hole statistical mechanics revealed
however that
an equilibrium state of a system of quantum black holes does
exist
and is describable with the use of the microcanonical
ensemble.

This fact alone suggests immediately the proper route for
quantization.
The proper vacuum state should belong to an energy
representation
where the sum of the particle energies is fixed by the
black hole
mass (for a static black hole). The vacuum state of the
theory is
thus parametrized by the black hole mass and is, formally,
the inverse
Laplace transform of the canonical vacuum. We shall call it
the
microcanonical vacuum. All correlation functions can in
principle
be calculated with such a vacuum and an expression for the
unperturbed
microcanonical propagator is obtained in section III. It is
essentially
Weldon's propagator\cite{wel}.

Stability of the microcanonical vacuum is insured by the
above
mentioned existence of a microcanonical equilibrium state
for quantum
black hole systems. By the bootstrap property, a gas of
black holes is
equivalent to a single quantum black hole and therefore we
expect the
microcanonical formulation to describe pure states. In
section III
arguments based on density matrix calculations corroborate
this
expectation.

In the following section, in order to make the paper self-
contained,
we first present a short review of the semiclassical
treatments of the
quantum black hole problem.

\section{Black Holes as P-Branes}

In Ref.\cite{hl1} we demonstrated that the model of black
holes as p-branes is free of the logical inconsistencies and
paradoxes of the thermodynamical interpretation
\cite{hawk1,bek} of black hole physics.  In this section we
review the semiclassical methods used to obtain information
from quantum gravity theory and discuss the rationale for
treating black holes as p-branes.

\subsection{The WKB Method}

The WKB method is one of the earliest nonperturbative
methods used in ordinary quantum mechanics.  In the path
integral formulation of quantum field theory the WKB method
amounts to finding the classical solutions to the Euclidean
equations of motion and then functionally integrating over
quantum fluctuations around these classical solutions.  The
Euclidean solutions, or instantons, describe the tunneling
of particles through the effective potential of the theory.

For a black hole the instanton solutions allow us to
calculate the semiclassical probability for a particle to
tunnel through the horizon.  The nature of the instanton
and the resulting expression for the Euclidean action depend
upon the charateristics of the black hole.  For example the
Euclidean spacetime for a $D$-dimensional Schwarzschild
black hole has a conical singularity, which is removed by
requiring that the imaginary time dimension be circular with
circumference $\beta_H$, the Hawking inverse
``temperature''.  The gravitational instantons in this case
are periodic instantons.  The Euclidean action determined
from the Euclidean metric
\begin{eqnarray}
ds^2 = e^{2\Phi} d\tau^2 + e^{-2\Phi}dr^2 + r^2 d\Omega_{D-
2}^2 \; ,
\end{eqnarray}
where
\begin{eqnarray}
e^{2\Phi} = 1 -\Bigl({r_+\over{r}}\Bigr)^{D-3} \; ,
\end{eqnarray}
and $r_+$ is the horizon radius, is given by
\begin{eqnarray}
S_E = {A_{D-2}\over{16\pi}}\beta_H r_+^{D-3} \; .
\end{eqnarray}
In this expression $A_D$ is the area of a unit $D$-sphere
and the vanishing of the conical singularity relates
$\beta_H$ to the horizon radius $r_+$
\begin{eqnarray}
\beta_H = {2\pi \over{[e^{\Phi}\partial_r
e^{\Phi}]_{r=r_+}}} = {4\pi r_+ \over{D-3}} \; .
\end{eqnarray}
The horizon radius is determined by the black hole mass $M$
\begin{eqnarray}
M = {(D-2)\over{16\pi}}A_{D-2} r_+^{D-3} \; .
\end{eqnarray}
In terms of the mass the Euclidean action can be written as
\begin{eqnarray}
S_E(M) ={\beta_H M \over{D-2}} = C(D) M^{{D-2\over{D-
3}}} \; ,
\end{eqnarray}
where $C(D)$ is defined as,
\begin{eqnarray}
C(D) = {4^{{D-1\over{D-3}}}\pi^{{D-2\over{D-3}}}
\over{(D-3)(D-2)^{{D-2\over{D-3}}} A_{D-2}^{{1\over{D-3}}}}}
\; .
\end{eqnarray}

The tunneling probability of a single particle escaping the
black hole in the WKB approximation,
\begin{eqnarray}
P \sim e^{-S_E(M)/\hbar} \; ,
\end{eqnarray}
is essentially the inverse of the quantum degeneracy of
states, so
\begin{eqnarray}
\sigma \sim e^{C(D) M^{{D-2\over{D-3}}}} \; .
\end{eqnarray}
This expression for the degeneracy of states when compared
to those of known non-local quantum theories shows that the
$D$-dimensional Schwarzschild black holes are quantum
excitation modes of a ${D-2\over{D-4}}$-brane.  The
implication of this result is that black holes are
elementary particles.

We have studied \cite{hl1,hl2,hl3,hl4} a gas of such
particles in the microcanonical ensemble and have shown that
they form a conformal theory in the sense that this ensemble
obeys the
statistical bootstrap condition and that the S-matrix is
dual.  For a gas of $N$ black holes the equilibrium state is
the one for which there is one very massive black hole and
$(N-1)$ massless black holes.

\subsection{Mean-Field Theory}

In this approximation \cite{birr} the S-matrix elements are
calculated using $in$ and $out$ states which are
assumed to be free in the distant past and distant future.
In the present case we wish to consider quantum fields
scattering off the black hole horizon.  There is a doubling
of the number of degrees of freedom in this case because the
horizon divides space into two causally disconnected
regions, requiring two different Fock spaces.  The
mathematical structure used to describe this situation is
the
same as that for field theories at finite temperature, e.g.
the thermofield dynamics formalism\cite{umez}.  The states
describing the system are direct products of the basis
vectors of the Fock spaces of the two disconnected regions.
For example, the thermal vacuum state for outgoing particles
can be
written as
\begin{eqnarray}
|out,0> = Z^{-1/2}(\beta)\sum_{n=0}^{\infty} e^{-\beta
n\omega/2} |n> \otimes |\tilde{n}> \; ,
\end{eqnarray}
where $|n>$ and $|\tilde{n}>$ are the Fock spaces of the two
disconnected regions.  An observer outside the horizon sees
only the $|n>$ states directly.  The partition function
$Z(\beta)$ is given by
\begin{eqnarray}
Z = \sum_{n=0}^{\infty} e^{-\beta n \omega} \; ,
\end{eqnarray}
and for any observable operator $\cal O$ the vacuum
expectation value is
\begin{eqnarray}
<out,0|{\cal O}|out,0> = {1\over{Z(\beta)}}
\sum_{n=0}^{\infty} e^{-\beta n\omega} <n|{\cal O}|n> \; .
\end{eqnarray}
The inverse temperature $\beta$ in this case is determined
by the surface gravity $\kappa$ of the black hole
\begin{eqnarray}
\beta = {2\pi\over{\kappa}}
\end{eqnarray}
This is the same as the expression for $\beta_H$, which we
encountered in the WKB approximation.

If we take ${\cal O}$ in Eq.(2.12) to be the number
operator,
the particle number density for a given mass $m$ is
\begin{eqnarray}
n_k(m;\beta_H) = {1\over{e^{\beta_H \omega_k(m)} - 1}} \; .
\end{eqnarray}
In a local field theory, this
expression would present a problem because it implies a loss
of quantum coherence in the scattering process.  The $in$
state is a
pure state
\begin{eqnarray}
|in,0> = |0>\otimes |\tilde{0}> \; ,
\end{eqnarray}
but the expression in Eq.(2.14) clearly arises from a
mixture
of states, resulting in a failure of the unitarity principle
during the scattering process
\begin{eqnarray}
|out,0> = S^{-1}(\beta)|in,0> \; \ \ S^{-1} \neq S^{\dagger}
\; .
\end{eqnarray}

Since black holes are, in our point of view, p-brane quantum
excitations, Eq.(2.14) cannot be the final result.  Quantum
non-local effects (back
reactions) must be taken into account.  The expression
obtained from the thermal vacuum for a single mass state
must be summed over in order to take into account all
possible mass states
\begin{eqnarray}
n_k(\beta_H) = \int_0^{\infty} dm \sigma(m)\; n_k(m;\beta_H)
\; .
\end{eqnarray}
The expressions for the thermal vacuum and the canonical
partition function are then given by
\begin{eqnarray}
|out,0> = Z^{-1/2}(\beta)\bigl[\prod_{m,k}\ \
\sum_{n_{k,m}=0}^{\infty}\bigr] \prod_{m,k} e^{-
{\beta\over{2}} n_{k,m} \omega_{k,m}} |n_{k,m}>\otimes
|\tilde{n}_{k,m}> \; ,
\end{eqnarray}
and
\begin{eqnarray}
Z(\beta_H) = \exp\Bigl( -{V\over{(2\pi)^{D-1}}}\int_{-
\infty}^{\infty} d^{D-1}{\vec k} \int_0^{\infty} dm\sigma(m)
\ln[1-e^{-\beta_H\omega_k(m)}]\Bigr) \; ,
\end{eqnarray}
where $\omega_k(m) = \sqrt{\vec{k}^2 + m^2}$.  The partition
function can also be written in terms of the statistical
mechanical density of states
\begin{eqnarray}
Z(\beta_H) = \int_0^{\infty} dE e^{-\beta_H E} \Omega(E)\; .
\end{eqnarray}
Equating the latter expression to that in Eq.(2.19), we
obtain the self-consistency condition first written down by
Hagedorn\cite{hage} as a model of strong interactions at
high energy.  $\Omega(E)$ depends upon $\sigma(E)$, and we
have a statistical bootstrap requirement, which can be
stated as
\begin{eqnarray}
{\sigma(E)\over{\Omega(E)}} \to 1 \; ; \ \ (E \to \infty) \;
{}.
\end{eqnarray}
The unique solution of Hagedorn's condition together with
the bootstrap requirement is of the form
\begin{eqnarray}
\sigma(m) \sim e^{bm} \; ; \ \ (m \to \infty) \; ,
\end{eqnarray}
with the restriction that $\beta_H > b$, where $b^{-1}$ is
the Hagedorn temperature.  This expression for $\rho(m)$ is
the same as that obtained from string theories.

Our analysis of the tunneling probability in the WKB
approximation for a particle to escape a Schwarzschild black
hole showed that black holes are p-branes with $p ={D-
2\over{D-4}}$ (see Eq.(2.9)).  Since $p>1$ unless $D \to
\infty$,
black holes do not satisfy the self-consistency condition
because of the divergence of the canonical
partition function.  The black hole system is not in thermal
equilibrium, therefore there is no self-consistent solution
for the quantum density of states under the assumption of
thermal equilibrium.  The thermal vacuum is the false vacuum
for this system.

In order to properly quantize fields in black hole
backgrounds we must start from the true vacuum.  In the next
section we lay the foundation for the solution of this
problem by starting from the proper microcanonical
description and developing a relatively simple
expression for the true vacuum.
\narrowtext

\section{The Microcanonical Formulation}

The review of the preceding section clearly exhibited the
fact that,
taking into account the full non-locality of quantum gravity
theories
(p-brane theories), the ``thermal vacuum'' of the
traditional canonical
quantization of fields in black hole spacetimes is not
stable. It is
the false vacuum. In this section, with the help of our
knowledge
of black hole statistical mechanics, we shall find the true
vacuum
of the theory as well as an expression for the free particle
propagator, the basic object in perturbation theory.

To make rapid progress, let us re-express the canonical
(thermal)
vacuum (2.18) as follows,
\begin{eqnarray}
|O(\beta)>\;=\;{{\hat \rho}^{1\over 2}}({\beta};{\bf
H})\,|{\Im}>\;,
\end{eqnarray}
where ${\hat \rho(\beta)}$ is the normalized canonical
density
matrix operator
(acting solely on the $|n>$ subspace). It is given by,
\begin{eqnarray}
{\hat \rho}({\beta},{\bf H})\;=\;{\rho(\beta,{\bf
H})\over{
<\Im|\rho(\beta,{\bf H})|\Im>}} \; ,
\end{eqnarray}
where,
\begin{eqnarray}
{\rho}({\beta},{\bf H})\;=\;e^{-{\beta}{\bf H}}\;,
\end{eqnarray}
in which {\bf H} is the Hamiltonian operator and,
\begin{eqnarray}
|{\Im}>\;=\;\prod_{k,m} \sum_{n_{k,m}} |n_{k,m}>\otimes
|{\tilde n}_{k,m}>\;.
\end{eqnarray}
Notice that,
\begin{eqnarray}
{\rm tr}{\cal O}\;=\;<{\Im}|{\cal O}|{\Im}>\;,
\end{eqnarray}
for any operator ${\cal O}$ acting on the physical subspace
(as
seen
by the observer outside the horizon).

If we introduce the so-called thermal doublet notation,
\begin{eqnarray}
{\phi}^{a}\;=\;\left( \begin{array}{c} {\phi} \\
{{\tilde \phi}^{\dagger}} \end{array} \right)\;,
\end{eqnarray}
the free particle causal propagator is now given as,
\begin{eqnarray}
-i<{\Im}|T{{\hat \rho}^{1-\alpha}}(\beta){\phi^a}(x_1)
{\phi^b}(x_2){{\hat \rho}^{\alpha}}(\beta)|{\Im}>\;
=\;{{\Delta}_{\beta,\alpha}^{ab}}(x_1,x_2)\;.
\end{eqnarray}
At equilibrium, the choice of the parameter $\alpha$ is
completely
arbitrary. In other words, physical observables are
$\alpha$-independent.
This situation changes out of equilibrium however, but one
need not be
concerned by that here. The thermal vacuum  of Eq.(2.18)
corresponds
to the choice ${\alpha}={1 \over 2}$. The Fourier transform
of the
propagator (3.7) is now given as follows,
\begin{eqnarray}
\Delta_{\beta,\alpha}^{ab}(k) = {\tau_3\over{k^2+m^2-
i\epsilon\tau_3}} + {2\pi i
\delta(k^2+m^2)\over{e^{\beta|k_0|} - 1}}\left(
\begin{array}{cc}1 & e^{\alpha\beta|k_0|}\\
e^{(1-\alpha)\beta|k_0|} & 1 \end{array}
\right)
\end{eqnarray}
where ${\tau}_3$ is the Pauli matrix in thermal doublet
space.

For convenience, we now choose $\alpha=1$. So all connected
correlation
functions are calculated as follows,
\begin{eqnarray}
G_{\beta}^{{a_1}{a_2}\cdots{a_N}}(1,2,\cdots,N)\;=\;-
i<{\Im}|T
{\phi}^{a_1}(1){\phi}^{a_2}(2)\cdots{\phi}^{a_N}(N)|{\beta}>
\: ,
\end{eqnarray}
where,
\begin{eqnarray}
|\beta>\;=\;{\hat \rho}(\beta,{\bf H})|{\Im}>\;.
\end{eqnarray}
Our previous work on black hole statistical mechanics made
it clear
that the canonical ensemble doesn't exist for black holes
and that
the equilibrium state must be analyzed in the fundamental
microcanonical
ensemble. This piece of knowledge now strongly suggests the
proper
route for solving the present problem. We are then led to
define
(formally) the microcanonical vacuum $|E>$ as follows,
\begin{eqnarray}
|\beta>\;\equiv\;\int_{o}^{\infty}dE\,e^{-\beta E}|E>\;.
\end{eqnarray}
Physical correlation functions are now defined as follows,
\begin{eqnarray}
G_E^{{a_1}{a_2}\cdots{a_N}}(1,2,\cdots,N)\;=\;<{\Im}|T
{\phi}^{a_1}(1){\phi}^{a_2}(2)\cdots{\phi}^{a_N}(N)|E>\;.
\end{eqnarray}
Of course only those correlations with
${a_1}={a_2}=\cdots={a_N}=1$
are directly observable outside the black hole horizon.

Now since $<{\Im}|\beta>=1$, the microcanonical vacuum is
normalized as follows,
\begin{eqnarray}
<{\Im}|E>\;=\;{\delta}(E)\;.
\end{eqnarray}
Recalling that,
\begin{eqnarray}
Z^{\pm 1}(\beta,V)\;=\;\int_{o}^{\infty}dE\,e^{-\beta
E}{\Omega(E,\pm V)}
\;,
\end{eqnarray}
and,
\begin{eqnarray}
\rho(\beta,{\bf H})\;=\;\int_{o}^{\infty}dE\,e^{-\beta E}
\rho(E,{\bf H})\;,
\end{eqnarray}
where,
\begin{eqnarray}
\rho(E,{\bf H})\;=\;\delta(E-{\bf H})\;,
\end{eqnarray}
is the (un-normalized) microcanonical density matrix, and
also making use
of Eq.(3.10), we arrive at the following expression for the
microcanonical vacuum $|E>$,
\begin{eqnarray}
|E>\;=\;\Omega(E-{\bf H},-V)|{\Im}>\;,
\end{eqnarray}
in which the density of states $\Omega(E,V)$ is given as,
\begin{eqnarray}
\Omega(E,V)\;=\;\delta(E)\;+\;&&\sum_{n=1}^{\infty}
\Bigl[{V\over{(2 \pi)^{D-1}}}\Bigr]^{n}
{1\over{n!}}\Bigl[\prod_{i=1}^{n}\int_o^{\infty}d{m_i}
\sigma({m_i})\int_{-\infty}^{\infty}{d^{D-1}}{\vec k}_{i}
\sum_{{l_i}=1}^{\infty}\Bigr] \nonumber \\
&&\times\,{1\over{{l_1}{l_2}\cdots{l_n}}}\delta(E\,-
\,{\sum_{i=1}^n}
{l_i}{\omega_{k_i}({m_i})})\;,
\end{eqnarray}
where $\sigma(m)$ is the quantum black hole degeneracy of
states and,
\begin{eqnarray}
{\omega_k}(m)\;=\;\sqrt{{{\vec k}^2}\,+\,{m^2}}\;.
\end{eqnarray}
It may be useful to define a ``normalized'' microcanonical
vacuum as
follows,
\begin{eqnarray}
|O(E)>\;\equiv\;{{|E>}\over{\delta(0)}}\;,
\end{eqnarray}
so that,
\begin{eqnarray}
<{\Im}|O(E)>\;=\;1\;\;(E=0)\;;.
\end{eqnarray}
The set of Eqs. (3.12), (3.13), (3.16)-(3.20) therefore
properly
describes our quantum theory of fields in black hole
spacetimes.
Of course, we neglected here the interaction effects.
These can be taken into account in perturbation theory with
the
use of the following microcanonical causal propagator,
\FL
\begin{eqnarray}
\Delta_{E,1}^{ab} &=&
{1\over{\delta(0)}}\Bigl({\tau_3\delta(E)\over{k^2+m^2-
i\epsilon\tau_3}} + 2\pi i\delta(k^2+m^2)\nonumber \\
&\times& \biggl[\sum_{l=1}^{\infty} \delta(E-l|k_0|)
\left( \begin{array}{cc} 1 & \ \ 1 \\
1 & \ \ 1 \end{array}\right) + \delta(E)\left(
\begin{array}{cc}
0 & \ \ 1\\ 0 & \ \ 0 \end{array} \right)\biggr]\Bigr)
\end{eqnarray}
in which use has been made of Eq.(3.21).

The physical (1,1)-component of the above propagator matrix
is essentially Weldon's propagator.

As opposed to the thermal propagator of Eq.(3.8), the above
microcanonical propagator yields a particle number density
(Hawking radiation) seemingly describing that of a pure
state,
\begin{eqnarray}
n_{k,m}\;=\;\sum_{l=1}^{\infty}\,{{\delta(E\,-
\,l{\omega_k(m)})}
\over{\delta(0)}}\; .
\end{eqnarray}
A calculation supportive of this argument is based on the
following
expression for the ``normalized'' microcanonical density of
states
${\hat \rho}(E,{\bf H})$,
\begin{eqnarray}
{\hat\rho}(E,{\bf H})\;=\;{{\delta(E\,-\,{\bf
H})}\over{\delta(0)}}\;.
\end{eqnarray}
Therefore we have,
\begin{eqnarray}
{{\hat \rho}_{{E_1}{E_2}}(E)}\;=\;{{\delta(E-
{E_1})\delta({E_1}-
{E_2})}\over{\delta(0)}}\;,
\end{eqnarray}
and,
\begin{eqnarray}
\int_o^{\infty}dE\,{\hat \rho}_{EE}\;=\;1\;.
\end{eqnarray}

It is well known that a necessary and sufficient condition
for a
density matrix to describe a pure state is the following
idempotency condition,
\begin{eqnarray}
\int_o^{\infty}d{E'}\,{{\hat \rho}_{{E_1}{E'}}}{{\hat
\rho}_{{E'}
{E_2}}}\;=\;{{\hat \rho}_{{E_1}{E_2}}}\;.
\end{eqnarray}
It is immediate to show that the form (3.25) for the density
matrix does satisfy the requirement (3.28). The eigenvalue
of the
density matrix (3.25) is 1.

Of course, in the usual context of ordinary statistical
mechanics,
the form (3.24) for the microcanonical density matrix is an
idealization, an expression valid only approximatively at
macroscopic scales such that the mesoscopic statistical
fluctuations ${\delta}E$
are large compared to the quantum fluctuations ${\Delta}E$
(${{\delta}E}\gg{{\Delta}E}$), but nevertheless small when
seen
from the macroscopic viewpoint at which statistical
mechanics
describes the system. In this way the microcanonical
ensemble
still describes mixed states.

The situation here is quite different as there is no
statistical
fluctuation concept to start with. The present system is a
{\it bona fide} quantum system embedded in a classical
background.
There is no ensemble theory here. Consequently, the form
(3.24)
for the density matrix is {\bf not} an idealization but a
precise
statement, valid right down to the level of the quantum
fluctuations.
In this sense, the present microcanonical description is in
a pure
state.

\narrowtext
\section{Conclusion}

In this work, we presented a consistent quantization
scheme for field theories in black hole spacetime
backgrounds.

Results from the study of black hole statistical mechanics
strongly suggested a fixed energy (static black hole mass)
representation basis for the Hilbert space of the theory,
instead
of the usual ``thermal state''. Such a representation is
formally
constructed by taking the inverse Laplace transform of the
thermal
description. Of course, because of the form for the quantum
black hole degeneracy of states,
only the microcanonical (E-) representation
is well defined and leads to a stable vacuum.

We have further argued that, in the microcanonical
representation,
particle states in the black hole background (Hawking's
radiation)
are pure states, unlike the traditional thermal description.
This
conclusion was reached on the basis of the idempotency
condition
obeyed by pure states density matrices.

Of course, ever deeper understanding of the results
presented here will
be at the core of our future endeavors.

\acknowledgments

This research was supported in part by the U.S. Department
of Energy under Grant No. DE-FG05-84ER40141.

\end{document}